\documentclass[%
 reprint,
 amsmath,amssymb,
 aps,
]{revtex4-2}

\usepackage{graphicx}
\usepackage{dcolumn}
\usepackage{bm}
\usepackage{xcolor}
\usepackage{filecontents}
\usepackage{ulem}

\begin{document}

\preprint{APS/123-QED}

\title[Modification of the Optical Properties of Molecular Chains\\ upon Coupling to Adatoms]{Modification of the Optical Properties of Molecular Chains\\ upon Coupling to Adatoms}


\author{Marvin M. M\"uller}
\email{marvin.mueller@kit.edu}
\affiliation{Institute of Theoretical Solid State Physics, Karlsruhe Institute of Technology (KIT), 76131 Karlsruhe, Germany}
\author{Miriam Kosik}
\email{mkosik@doktorant.umk.pl}
\affiliation{Institute of Physics, Nicolaus Copernicus University in Toru\'n, Grudziadzka 5, 87-100 Toru\'n, Poland}
\author{Marta Pelc}
\affiliation{Institute of Physics, Nicolaus Copernicus University in Toru\'n, Grudziadzka 5, 87-100 Toru\'n, Poland}
\author{Garnett W. Bryant}
\affiliation{Joint Quantum Institute, University of Maryland and National Institute of Standards and Technology, College Park, Maryland 20742, USA}
\affiliation{Nanoscale Device Characterization Division, National Institute of Standards and Technology, Gaithersburg, Maryland 20899, USA}
\author{Andr\'es Ayuela}
\affiliation{Donostia International Physics Center (DIPC), Paseo Manuel Lardizabal 4, 20018 Donostia-San Sebasti\'an, Spain}
\affiliation{Centro de F\'isica de Materiales, CFM-MPC CSIC-UPV/EHU, Paseo Manuel Lardizabal 5, 20018 Donostia-San Sebasti\'an, Spain}
\author{Carsten Rockstuhl}
\affiliation{Institute of Theoretical Solid State Physics, Karlsruhe Institute of Technology (KIT), 76131 Karlsruhe, Germany}
\affiliation{Institute of Nanotechnology, Karlsruhe Institute of Technology (KIT), 76021 Karlsruhe, Germany}
\author{Karolina S\l{}owik}
\affiliation{Institute of Physics, Nicolaus Copernicus University in Toru\'n, Grudziadzka 5, 87-100 Toru\'n, Poland}


\begin{abstract}
Adsorbed atoms (adatoms) coupled to the matrix of solid state host materials as impurities can significantly modify their properties. Especially in low-dimensional materials, such as one-dimensional organic polymer chains or quasi-one-dimensional graphene nanoribbons, intriguing manipulation of the optical properties, such as the absorption cross section, is possible. The most widely used approach to couple quantum emitters to optical antennas is based on the Purcell effect. This formalism, however, does not comprise charge transfer from the emitter to the antenna, but only spontaneous emission of the quantum emitter into the tailored photonic environment, that is evoked by the antenna. To capture such effects, we present a tight-binding formalism to couple an adatom to a finite Su-Schrieffer-Heeger chain, where the former is treated as a two-level system and the latter acts as an optical antenna. We systematically analyze how the coupling strength and the position of the adatom influence the optical properties of the molecular chains in the model. We take into account charge transfer from the adatom to the antenna and vice versa via an inter-system hopping parameter, and also include Coulomb interaction within the antenna as well as between the adatom and the antenna. We show that coupling the adatom to one of the bulk atoms of the linear chain results in a substantial change in optical properties already for comparatively small
coupling strengths. We also find that the position of the adatom crucially determines if and how the optical properties of the chains are altered. Therefore, we identify this adatom-chain hybrid system as a tunable platform for light-matter interaction at the nanoscale.
\end{abstract}

\maketitle


\section{\label{sec:introduction} Introduction}
The Su-Schrieffer-Heeger (SSH) model constitutes a simple yet powerful and instructive tight-binding (TB) based model to describe the electronic and topological properties of solids, and induced a large body of literature within the last four decades~\cite{Su1979, Kivelson1982, Fradkin1983, Harigaya1992, Quemerais1993, Atala2013, Ando2013, Meier2016, Asboth2016, DiLiberto2016, Turker2018, Lieu2018, Yuce2019}. Besides being a playground to explore topological phases~\cite{Atala2013, Yang2018} and quasiparticles~\cite{Su1979, Heeger1988, Nadj2014, Feldman2017}, it is also capable of revealing transport properties of organic polymers such as polyacetylene~\cite{Heeger2001} and the electronic energy level diagrams of molecular chains, for instance, if applied to finite systems. Moreover, it is to a large extent analytically solvable and, therefore, allows for powerful conceptual insights into the underlying physical principles. Within the model, it can be readily decided if a given atomic chain is electronically conducting or acts as an insulator. Additionally, non-trivial topological phases and the appearance of near-zero energy edge states in finite chains can be investigated. All these features can be traced back to chains of atoms with only slightly different coupling constants that give rise to three fundamentally different systems: the linear atomic chain (conductor), the dimerized atomic chain (conventional insulator), and the topological insulator.\\
These systems are physically realized in nature through several organic molecules. Linear polyenes, for instance, exhibit electrons that occupy $p_z$-orbitals of their hosting carbon atoms. Therefore, these molecules can be understood as a realization of a one-dimensional (1D) electron gas that is almost completely delocalized along the molecule. Hence, an analogy to the \textit{metallic} homogeneous 1D electron gas may be established~\cite{Bernadotte2013}. Polyacetylenes, on the other hand, constitute representatives of \textit{insulating} organic polymers with alternating bond strengths between neighboring carbon atoms and only exhibit considerable electronic transport upon doping~\cite{Park1980, Heeger1981, Basescu1987}. Moreover, 1D atomic chains can as well be realized artificially by growing them on a substrate \cite{Yanson1998, Takai2001, Ayuela2002, Nilius2002, Folsch2004, Blumenstein2011, Celotta2014, Salfi2016}. Quasi-1D reconstructions formed by metal deposition on silicon wafers~\cite{Yeom1999, Riikonen2006}, and also vacancies with dangling bonds on silicon wafers may form 1D chains of atom-like systems~\cite{Haider2009, Schofield2013, Kepenekian2014, Wyrick2018}.\\
We want to investigate the effect of introducing an electronically coupled adsorbed atom (adatom) into the TB description of the above mentioned finite SSH chains.
Coupling the adatom may result in symmetry breaking of the structure leading to a modification of its electronic properties and optical response \cite{Buchs2021}. Here, we aim to identify the parameters in the system that - if altered - influence these properties most sensitively. To derive statements that are as general as possible, we choose to couple the adatom to the SSH chains, \textit{i.e.}, generic models that represent \textit{metallic} and \textit{insulating} systems, and an insulating system that additionally hosts near-zero energy edge states. Our study reveals that the metallic linear chain is most prone to considerable modifications of its optical properties among the mentioned SSH chains upon coupling to an adatom.\\
Structural modifications of carbon nanostructures can be intrinsic, in the form of carbon adatoms 
\cite{Lehtinen2003} or lattice defects~\cite{Pedersen2008, Fuerst2009}, or extrinsic, given by foreign adatoms \cite{Chan2008, Ishii2008, Sanchez2009}. In particular, transition metal adatoms interact with carbon nanostructures through their $p_z$-orbitals \cite{Alonso2016, Alonso2017, Uchoa2008, Santos2010} and, therefore, can be described within the localized $\pi$-electron picture. The adatoms can be treated with the extended H\"uckel molecular orbital model~\cite{Ammeter1977, Ihnatsenka2011}, a nonmagnetic counterpart of the Anderson impurity model~\cite{Anderson1961, Sollie1991} that has been successfully employed before by the group of Jaroslav Fabian~\cite{Gmitra2013, Irmer2015, Zollner2016, Frank2017, Irmer2018}. They coupled one-level adatoms to extended bulk graphene in a TB framework to investigate spin-orbit coupling. In our approach, we treat the adatom as an effective two-level system (TLS), allowing for intra-impurity charge dynamics, such as spontaneous emission from the excited to the ground state of the adatom. The adatom is coupled consecutively to different atomic sites of the chain at various coupling strengths.\\
The article is structured as follows: We first introduce the model Hamiltonian, three scalar measures to characterize the single-particle eigenstates of said Hamiltonian, and discuss the optical properties that follow from it in Sec.~\ref{sec:theory}. In Sec.~\ref{sec:SSHmodel}, we investigate the stand-alone SSH chains without the adatom as a reminding preparatory work. Section~\ref{sec:hybridsystems} focuses on the hybrid chain-adatom systems without Coulomb interaction, whereas in Sec.~\ref{sec:interactingsystem}, we take into account electron-electron interaction before we summarize our findings in Sec.~\ref{sec:summary}. 

\section{\label{sec:theory} Theory}
We assume the hybrid system to consist of two components: a nanoscopic SSH chain that acts as an optical antenna and an adatom which is effectively described as a TLS. Both the adatom and the antenna are treated in a TB framework. In particular, we assume one mobile electron per carbon atom in the antenna's $p_z$-orbitals $\{|l\rangle\}$ that are localized at $\mathbf{r}_l$ in the vicinity of the corresponding host atoms $l\in[1,N_a]$ for an antenna of $N_a$ atoms. Mediated through $\pi$-bonds that connect $p_z$-orbitals of neighboring atoms $l$ and $l'$, electrons may change their location with a probability quantified by the TB hopping parameters $t_{ll'}$. They are proportional to the overlap integral of neighboring $p_z$-orbitals. For simplicity, we do not take into account the spin degree of freedom.\\
The TLS is characterized by its ground and excited states $|g\rangle$ and $|e\rangle$, representing two active orbitals of the adsorbed impurity with energies fixed at $E_g=-0.5\,\rm eV$ and $E_e=0.5\,\rm eV$ relative to the isolated antenna's energy levels. The adatom is coupled to one of the antenna's carbon atoms only, as it is the case for hydrogen, fluorine, and hydroxyl groups as adatoms, for instance~\cite{Ihnatsenka2011, Santos2012}. The ground and excited states couple to the host atom in the antenna via hopping parameters $t_g$ and $t_e$.

\subsection{\label{sec:system} Model Hamiltonian}
The system Hamiltonian consequently reads
\begin{align}
    H &= H_{\rm antenna} + H_{\rm TLS} + H_{\rm interaction}\nonumber\\
    &= -\sum_{l<l', \langle l,l'\rangle}t_{ll'}\Big(|l\rangle\langle l'| + |l'\rangle\langle l|\Big) \nonumber\\
    &\quad + E_e|e\rangle\langle e|+ E_g|g\rangle\langle g|\nonumber\\
    &\quad + t_e\Big(|l_c\rangle\langle e| + |e\rangle\langle l_c|\Big) + t_g\Big(|l_c\rangle\langle g| + |g\rangle\langle l_c|\Big),\label{eq:Hsystem}
\end{align}
where the atomic site indices $l,l'$ run over the antenna atoms, $\langle l,l'\rangle$ denotes a pair of nearest neighbor atoms, and $l_c$ is the antenna's atomic site the adatom is coupled to. We denote the $N=N_a+2$ energy eigenstates of the hybrid system by $\{|j\rangle\}$, where $j\in[1,N]$ and
\begin{align}
    H|j\rangle = E_j|j\rangle.
\end{align}
Here, the energies $E_j$ are given relative to the TB onsite energies which are set to zero in the Hamiltonian in Eq.~\eqref{eq:Hsystem}. The energy eigenstates may be expanded into the complete and orthonormal real-space atomic site basis $\{|l\rangle\}\cup \{|g\rangle, |e\rangle\}$ according to 
\begin{align}
    |j\rangle = c_{je}|e\rangle+c_{jg}|g\rangle + \sum_{l=1}^{N_a}c_{jl}|l\rangle.
\end{align}

\subsection{\label{sec:measures} State Characterization}
In the joint antenna-adatom system, the energy eigenstates of the
stand-alone antenna and the ones of the adatom hybridize, making it difficult to identify the stand-alone modes. In this paper, it is our goal to determine the conditions under which the presence of the adatom considerably modifies the optical properties of the isolated antenna. We especially focus on the tunability of the chain modes.
To achieve this goal, we put in place scalar measures for certain properties of the energy eigenstates that help to understand, quantify, and illustrate the changes that the electronic structures of the systems undergo. These measures map single-particle energy eigenstates to real numbers and, therefore, provide an intuitive manner to quantify their characteristics, which translate into optical properties, and to assign a physically meaningful order to them. The measures characterize the hybrid system for different parameter sets and especially for various coupling strengths and coupling positions. For the sake of brevity, from now on we use $\{|\tilde{l}\rangle\}=\{|l\rangle\}\cup \{|g\rangle, |e\rangle\}$ for the set of all real-space based active orbitals in the hybrid system.

\subsubsection{\label{sec:localization} State Localization}
We introduce the \textit{localization} $L_{|j\rangle}$, a measure that quantifies how strongly state $|j\rangle$ is localized on certain antenna sites $|l\rangle$ or adatom orbitals $|e\rangle$ and $|g\rangle$,
\begin{align}
    L_{|j\rangle} = \frac{\left(1-p_{|j\rangle}\right)N}{N-1}\quad   \in[0, 1],
    \label{eq:localization}
\end{align}
where the participation ratio~\cite{Bell1970, Bell1972, Wimmer2010, Downing2021}
\begin{align}
    p_{|j\rangle} = \frac{\left(\sum_{\tilde{l}} |c_{j\tilde{l}}|^2\right)^2}{N\sum_{\tilde{l}} |c_{j\tilde{l}}|^4}=\left(N\sum_{\tilde{l}} |c_{j\tilde{l}}|^4\right)^{-1} \, \in[1/N, 1]
    \label{eq:participation_ratio}
\end{align}
is a measure for the number of atomic site orbitals $|\tilde{l}\rangle$ that are significantly involved in the spatial distribution of the energy eigenstate $|j\rangle$ (the second equality in Eq.~\ref{eq:participation_ratio} holds for normalized states only). If the spatial distribution of state $|j\rangle$ is uniform on all the sites in the system, \textit{i.e.} $c_{j\tilde{l}}=1/\sqrt{N}$, then $p_{|j\rangle}=1$ and $L_{|j\rangle}=0$, and we call the state completely \textit{delocalized}. For a state localized on a single site $l_0$, \textit{i.e.} $c_{j\tilde{l}}=\delta_{\tilde{l}l_0}$, we obtain $p_{|j\rangle}=1/N$ and $L_{|j\rangle}=1$, and consequently call the state \textit{fully localized}.

\subsubsection{\label{sec:hybridization} State Hybridization}
To measure how strongly an eigenstate of the stand-alone isolated antenna is disturbed and modified by the presence of the adatom, we introduce the \textit{hybridization} $h_{|j\rangle}$. It is defined as
\begin{align}
    h_{|j\rangle} = 1-|\langle j|j^0\rangle|,
\end{align}
where $|j^0\rangle$ is an energy eigenstate of the Hamiltonian Eq.~\eqref{eq:Hsystem} for $t_e=t_g=0$ that evolves to $|j\rangle$ when the coupling is turned on. Hence, in the completely decoupled system, we have $|j\rangle=|j^0\rangle$ and $h_{|j\rangle}=0\quad\forall j\in [1,N]$. We want to emphasize here, that this definition of hybridization depends on the order (index) of the states. Therefore, it is necessary to scan the spectrum for energy level crossings before interpreting the results.

\subsubsection{\label{sec:activity} State Activity}
We are particularly interested in the optical properties of the hybrid system and the modifications thereof as we increase the coupling strength and change the position of the adatom. Therefore, it is not only necessary to identify the configurations which modify the electronic states in general, but in particular we aim to modify the set of states that is \textit{optically active}, \textit{i.e.}, that is responsible for the optical properties. To quantify if and to what extent a state is involved in the optical interaction, \textit{i.e.}, how strongly it contributes to the optical absorption cross section of the system, we define the state activity $a_{|j\rangle}$ of state $|j\rangle$ as
\begin{align}
    a_{|j\rangle} = \sum_{j'=1}^N|s_{jj'}|,
\end{align}
where $s_{if}=\left|E_{f}-E_{i}\right||\langle f|\hat{\mathbf{r}}|i\rangle|^2$ is the oscillator strength of the electronic single-particle transition $|i\rangle\rightarrow |f\rangle$ with the real-space position operator $\hat{\mathbf{r}}$ acting as $\langle l'|\hat{\mathbf{r}}|l\rangle=\mathbf{r}_l\delta_{ll'}$. In case $a_{|j\rangle}\approx 0$, we call $|j\rangle$ \textit{optically inert}. For non-degenerate states, this is equivalent to vanishing transition dipole moments between state $|j\rangle$ and all other states $|j'\rangle$, for example for symmetry reasons. High state activities, on the other hand, identify the given state as a donor or acceptor state for single-particle transitions in the hybrid system.

\begin{figure*}
    \centering
    \includegraphics[width=1.0\textwidth]{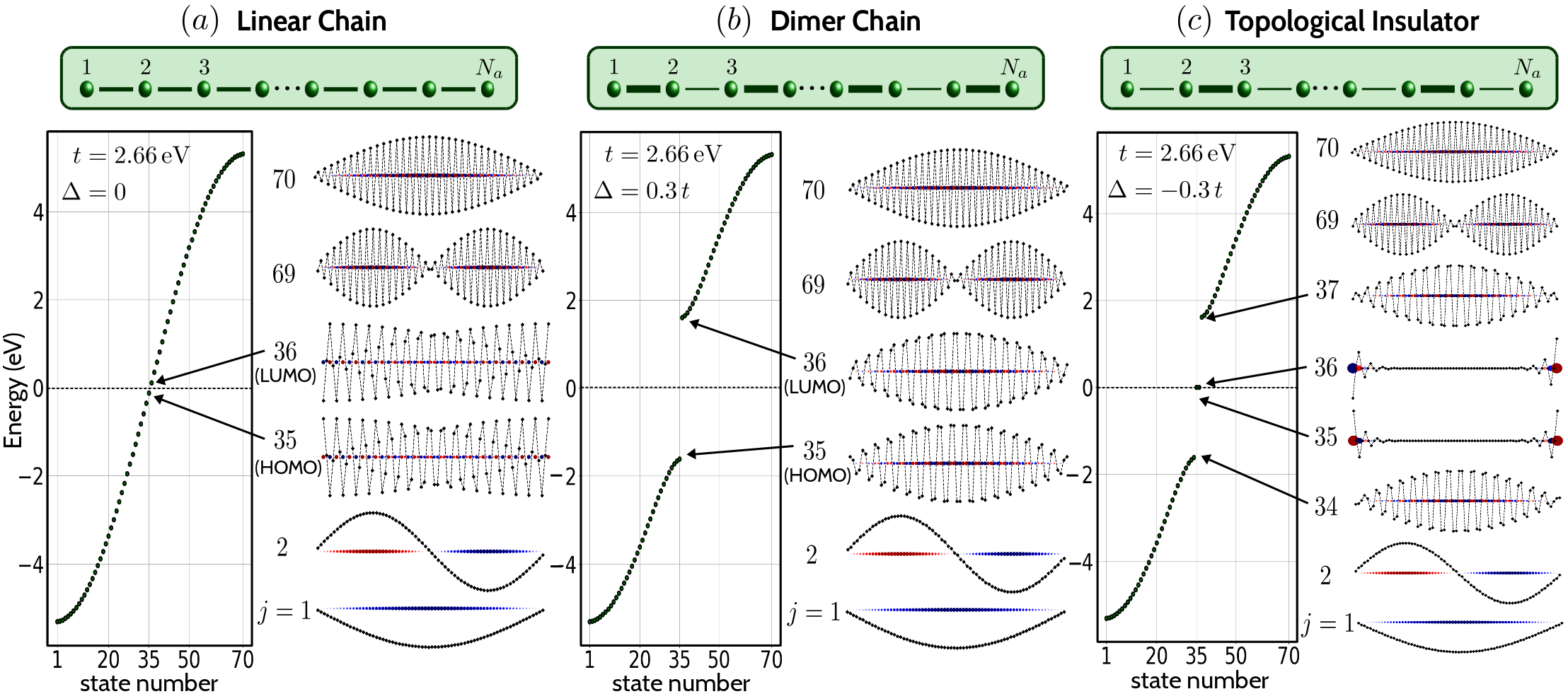}
    \caption{Jab\l{}onski energy level diagrams (bottom left panel) and real-space illustrations of single-particle states (bottom right panel) of (a) the linear chain, (b) the dimer chain, and (c) the topological insulator composed of $N_a=70$ atoms. The chains are illustrated in the top panel, where solid dark lines between neighboring atoms represent strong bonds and dashed light lines represent weak bonds. The black diamonds and color of the circles in the bottom right panels represent the real-valued expansion coefficients $c_{jl}$ of the states $|j\rangle$, whereas the size of the colored circles encodes their squared absolute values $|c_{jl}|^2$. We show the two single-particle states that are lowest and highest in energy, $j\in\{1, 2\}$ and $j\in\{69, 70\}$, respectively. They are qualitatively equivalent for all three structures. Moreover, we depict representatives of the states that are most relevant for the optical interaction of the structure. They are located around the particle-hole symmetry line at $E=0$, which is also the Fermi energy for half filling. The topological insulator exhibits two strongly localized (nearly) degenerate edge states inside the band gap close to $E=0$.}
    \label{fig:chains_with_states}
\end{figure*}

\subsection{\label{sec:optical_properties} Optical Properties}
As the central figure of merit to characterize the optical properties of the system we choose the linear absorption cross section $\sigma_{\rm abs}(\omega)$. All measures mentioned above are quantizers that characterize single-particle energy states. So far, we have not been asking if these states are actually occupied by electrons or not. This is, however, crucial to determine the absorption cross section, which makes it a property not only of the energy level diagram itself, but also of the number of electrons that populate it. Throughout the whole paper, we assume half filling of the energy landscape, corresponding to one mobile electron per atomic site orbital. Consequently, all states below the Fermi energy are occupied by two electrons and are unoccupied above. To isolate the interaction-mediated effects from the characteristics of the optical response that rely on the single-particle energy level diagram, we distinguish between the \textit{non-interacting} and the \textit{interacting} absorption cross sections, $\sigma_{\rm abs}^{\rm ni}(\omega)$ and $\sigma_{\rm abs}^{\rm i}(\omega)$. The latter includes Coulomb interaction between electrons in the system, whereas the former does not.\\
The \textit{non-interacting} absorption cross section of the hybrid system can be expressed as~\cite{Yamamoto2006}
\begin{align}
    \sigma_{\rm abs}^{\rm ni}(\omega) \propto \sum_{if} s_{if}\,\delta_\varepsilon\left( E_{f}-E_{i}-\hbar\omega \right),  
    \label{eq:noninteracting}
\end{align}
where $s_{if}$ is again the oscillator strength and $\delta_\varepsilon$ denotes Dirac's delta distribution broadened to a Lorentzian by a parameter $\varepsilon= 20\,\rm meV$ according to $ \delta_\varepsilon(x) = 2\varepsilon/(x^2+\varepsilon^2)$. The indices $i \in [1, j_{\rm HOMO}]$ and $f \in [j_{\rm LUMO}, N]$ denote the set of \textit{occupied initial} single-particle states from below the Fermi energy and \textit{unoccupied final} single-particle states from above the Fermi energy of the non-interacting system, respectively, that contribute to the transition $|i\rangle\rightarrow |f\rangle$. \\
To compute the \textit{interacting} absorption cross section $\sigma_{\rm abs}^{\rm i}(\omega)$, we probe the system with a small-amplitude spectrally broad electric field pulse $\mathbf{E}(t)=E(t)\mathbf{\hat{e}_x}$, polarized along the chain direction ($x$), and record the resulting dipole moment $\mathbf{p}(t)$. The system's response to a pulse polarized perpendicular to the chain direction is much smaller and at much higher energy and, therefore, neglected in this work. The way we take into account the induced Coulomb interactions and details on the computation of $\mathbf{p}(t)$ can be found in App.~\ref{app:time_dependent}. After Fourier transforming both quantities, we calculate the frequency-dependent polarizabilities according to $\alpha_{x,x}(\omega)=p_x(\omega)/E_x(\omega)$ and $\alpha_{x,y}(\omega)=p_y(\omega)/E_x(\omega)$. We then obtain the interacting absorption cross sections as
\begin{align}
\sigma_{x/y,\rm abs}^{\rm i}(\omega)\propto \omega\, \mathrm{Im}[\alpha_{x,x/y}(\omega)],
\label{eq:interacting}
\end{align}
and $\sigma_{\rm abs}(\omega)=\sigma_{x,\rm abs}(\omega)+\sigma_{y,\rm abs}(\omega)$, where $\mathrm{Im}[\cdot]$ denotes the imaginary part. In Eq.~\eqref{eq:interacting}, the Coulomb part is scaled by the parameter $\lambda$ that (numerically) controls the Coulomb interaction strength. Setting $\lambda$ to 0 retrieves the non-interacting absorption cross section in Eq.~\eqref{eq:noninteracting} (see details in App.~\ref{app:time_dependent}.).

\begin{figure}
    \centering
    \includegraphics[width=0.47\textwidth]{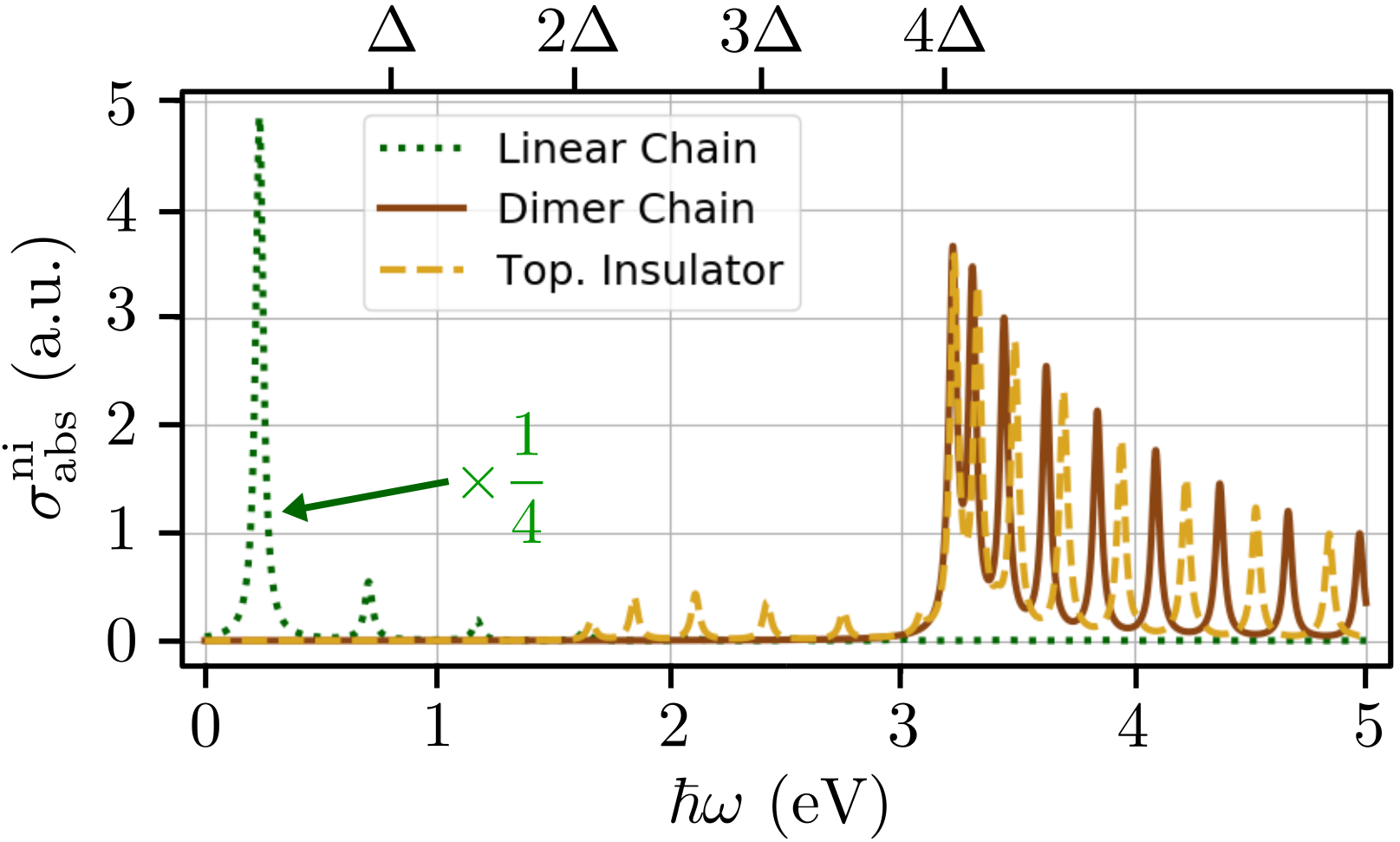}
    \caption{Non-interacting absorption spectrum $\sigma^{\rm ni}_{\rm abs}(\omega)$ for the linear chain (green dotted line), the dimer chain (brown solid line), and the topological insulator (yellow dashed line) for the parameter set given in Sec.~\ref{sec:SSHmodel} assuming half filling of the energy level diagram. The data of the linear chain have been scaled with the factor 1/4 to match the order of magnitude of the other two structures.}
    \label{fig:noninteracting_line}
\end{figure}

\section{Discussion and Results}
In the following, we successively discuss the electronic and optical properties of the stand-alone SSH chains, the hybrid chain-adatom system without Coulomb interaction, and finally the interacting hybrid chain-adatom system. 

\subsection{\label{sec:SSHmodel} Stand-Alone 1D SSH Chains}
As a first application, we study the three 1D molecular chains of the SSH model: the linear chain, the dimerized chain, and the topologically insulating chain. To create a topologically non-trivial system, we choose the number of atoms $N_a$ in our system to be even. The Hamiltonian reads
\begin{align}
    H_{\rm TB}^{\rm chains} &= -(t+\Delta)\sum_{\text{odd }l=1}^{N_a-1}\Big(|l\rangle\langle l+1| + |l+1\rangle\langle l|\Big)\nonumber\\
    &\quad -(t-\Delta)\sum_{\text{even }l=2}^{N_a-2}\Big(|l\rangle\langle l+1| + |l+1\rangle\langle l|\Big),
\end{align}
where we use the hopping parameter value of bulk graphene $t=t_{ll'}=2.66\,\rm eV$~\cite{Pelc2013} and $\Delta=0$ for the linear chain, $\Delta=0.3t$ for the dimer chain, and $\Delta=-0.3t$ for the topological insulator (see schematic plots in Fig.~\ref{fig:chains_with_states}). The transition from the semiconducting or insulating topologically trivial dimer chain $(\Delta>0)$ to the non-trivial topological insulator $(\Delta<0)$ takes place by crossing $\Delta=0$ via the gap-less linear chain. As $\Delta$ approaches zero from above, the dimer chain's band gap decreases, it closes for $\Delta=0$ (linear chain), and opens up again for negative $\Delta$, however, bringing forth the two near-zero edge states of the topological insulator.\\ 
Figure~\ref{fig:chains_with_states} shows the energy level diagrams and several selected characteristic single-particle states of (a) the linear chain, (b) the dimer chain, and (c) the topological insulator made up by $N_a=70$ atoms. As mentioned above, we assume half filling of the energy landscape, such that all states below (above) the Fermi energy $E=0$ are doubly occupied (unoccupied) in the linear chain and the dimer chain. The topological insulator exhibits two nearly degenerate edge states close to the Fermi energy $E=0$ that we populate with one electron each. We immediately notice that both the low-energy and the high-energy states are conceptually equivalent for all three structures. The physical difference between the systems becomes more pronounced the closer one gets to the energetic region around the Fermi energy $E=0$. However, this is also the energetic region where we find the single-particle states that are predominantly active in the optical interaction of the investigated systems. Therefore, we can indeed expect substantially differing optical responses from the three structures as we will show in the following.
\begin{figure}
    \centering
    \includegraphics[width=0.48\textwidth]{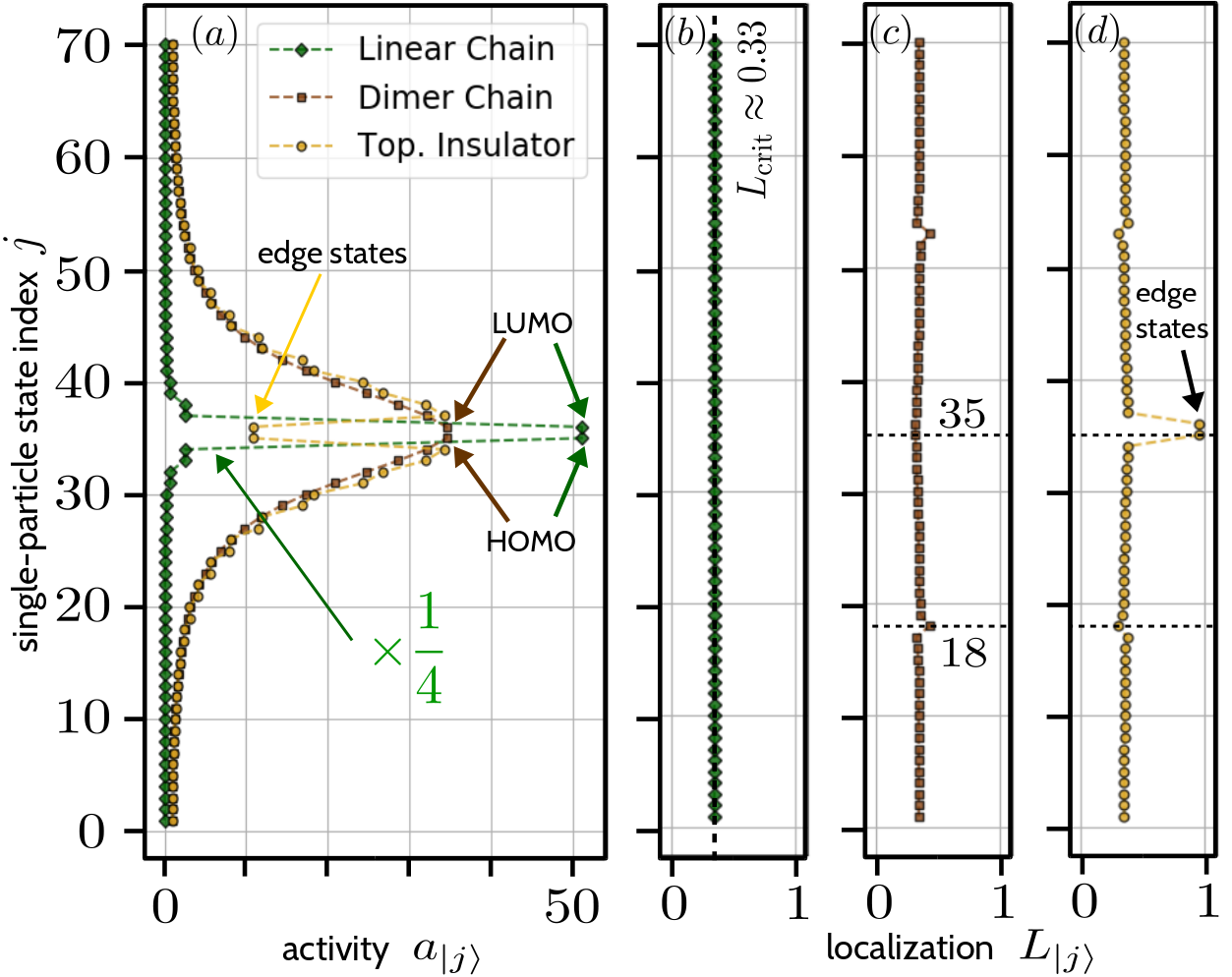}
    \caption{(a) State activity $a_{|j\rangle}$ of the three molecular chains of the SSH model. The data of the linear chain (green diamonds) have been scaled with the factor of 1/4 to match the order of magnitude of the other two structures. (b-d) State localizations $L_{|j\rangle}\in[0,1]$ of the single-particle states of the linear chain (b), the dimer chain (c), and the topological insulator (d). While the linear and dimer chains exhibit localization values around the critical value $L_{\rm crit}\approx 0.33$, the topological insulator's near-zero energy states $j=35$ and $j=36$ localize strongly at the edges of the chain, \textit{c.f.} Fig.~\ref{fig:chains_with_states}c, and nearly reach $L_{|j\rangle}\approx 1$.}
    \label{fig:activity_localization}
\end{figure}

\subsubsection{\label{sec:linear_chain} Linear Chain}
The discrete energy level diagram of the finite linear chain in Fig.~\ref{fig:chains_with_states}a results from quantizing the metallic band structure of the infinite chain. Additionally to the above mentioned low- and high-energy states, we show the HOMO and LUMO states. Their structures can be described as two nested modes of quarter wavelength shape of even and odd symmetry, respectively, on the two sublattices of the chain. In Fig.~\ref{fig:noninteracting_line}, the green dotted line shows the linear chain's non-interacting absorption cross section as a function of the excitation energy. The energy of the most prominent low-energy absorption mode around $\hbar\omega=0.24\,\rm eV$ coincides exactly with the energy difference of the HOMO and LUMO states. To confirm the obvious conclusion, we quantify the contributions of single-particle transitions in the linear chain to the absorption spectrum with the state activity $a_{|j\rangle}$. Figure~\ref{fig:activity_localization}a (green diamonds) shows the state activity of all single-particle states of the linear chain. Indeed, we note that the HOMO and LUMO are the only states that significantly contribute to the non-interacting absorption spectrum. It can, therefore, be concluded that the prominent low-energy mode at $\hbar\omega=0.24\,\rm eV$ corresponds to the electronic transition $|j_{\rm HOMO}\rangle\rightarrow |j_{\rm LUMO}\rangle$. Furthermore, we note, that besides the states HOMO$-2$, HOMO$-1$, LUMO$+1$, and LUMO$+2$, all other single-particle states are optically inert, \textit{i.e.}, $a_{|j\rangle}\approx 0$. This is in stark contrast to the activity of the single-particle states of the dimer chain and the topological insulator, as can be seen from Fig.~\ref{fig:activity_localization}a as well (brown squares and yellow circles). To engineer the optical properties of the linear chain, it is, therefore, desirable to either modify the optically active HOMO and LUMO states or to increase the optical activity of other states that are located further away from the Fermi energy by means of coupling the adatom to the system.\\
Figures~\ref{fig:activity_localization}b-d show the localizations $L_{|j\rangle}$ of the three chains' states. It is interesting to notice that the localization of all the linear chain's states have the exact same value and do not fall below a critical localization $L_{\rm crit}$. We can compute this value by plugging the analytical solution of the SSH model~\cite{Atala2013} for the linear chain's lowest-energy state $c_{1l}=\sqrt{\frac{2}{N_a+1}}\sin\left(\frac{\pi l}{N_a+1}\right)$, for instance, into Eqs.~\eqref{eq:participation_ratio} and \eqref{eq:localization}. We obtain $L^{\rm chain}_{|1\rangle}=\frac{1}{3}\left(1-\frac{1}{N_a-1}\right)$, which evaluates to 0.33 for a chain of length $N_a=70$. It can further be shown, that \textit{all} single-particle states of the linear chain evaluate to this exact same value, $L^{\rm chain}_{|j\rangle}=L_{\rm crit}$ independent of $j$.

\subsubsection{Dimer Chain}
The energy level diagram of the dimer chain in Fig.~\ref{fig:chains_with_states}b is of insulating character. We observe a lower-lying and a higher-lying quasi-continuum of states which would constitute the valence band and the conduction band in the limit of an infinitely extended chain $(N_a\rightarrow\infty)$, with a band gap of size $4|\Delta|\approx 3.19\,\rm eV$. Besides the low- and high-energy states, Fig.~\ref{fig:chains_with_states}b also shows the HOMO and LUMO states of the dimer chain. They display two nested modes of half wavelength shape of even and odd symmetry, respectively. We notice that especially the states around the Fermi energy exhibit a dimerized nature, \textit{i.e.}, neighboring atoms act alike and behave collectively as a two-atomic unit cell and not as individual atoms anymore. This is not the case for the low- and high-energy modes that conceptually look similar to the corresponding modes of the linear chain. Just as for the linear chain, the single-particle transition from HOMO to LUMO produces the most prominent resonance at $3.19\,\rm eV$ at the lower edge of the quasi-continuum in Fig.~\ref{fig:noninteracting_line} (brown solid line). Unlike in the case of the linear chain, however, the absorption spectrum is much richer. We observe many more modes above $3.19\,\rm eV$ that are of the same order of magnitude as the most prominent one. A way to consistently complement this finding is through the state activities of the dimer chain in Fig.~\ref{fig:activity_localization}a (brown squares). Although the HOMO and LUMO states exhibit the highest state activity here as well, many states around $E=0$ are optically active and contribute to the absorption spectrum, and none of them is completely inert. As a consequence, many pairs of optically active states couple and lead to the formation of the quasi-continuum of comparatively dense lying absorption modes above the band gap.\\
The slightly different localization values for the dimer chain's and topological insulator's states $j=18\approx N_a/4$ and $53\approx 3N_a/4$ with respect to other states in Figs.~\ref{fig:activity_localization}c and d do not affect the optical properties of the structure substantially due to the low activity of these states.

\subsubsection{Topological Insulator}
The energy level diagram of the topological insulator in Fig.~\ref{fig:chains_with_states}c strongly resembles the one of the dimer chain and is of insulating character as well. The states $j=34$ and $j=37$ are conceptually equivalent to the HOMO and LUMO of the dimer chain. However, we additionally find two near-zero degenerate states inside the band gap. They are strongly localized at the edges of the chain and attain localization values close to 1, as is shown in Fig.~\ref{fig:activity_localization}d. Moreover, Fig.~\ref{fig:activity_localization}a reveals that they are mildly optically active as well, which leads to the formation of a few absorption peaks in Fig.~\ref{fig:noninteracting_line} (yellow dashed line) on the outskirts of the quasi-continuum in the range between $\hbar\omega=2|\Delta|$ and $\hbar\omega=4|\Delta|$. This distinguishes the absorption spectrum of the topological insulator from the one of the dimer chain of equal length. While the most prominent mode at $3.19\,\rm eV$ is present in both insulating systems, the spectral position of the modes differ more the higher the energies of the modes get. The highest energy modes of the dimer chain and topological insulator, presented in Fig.~\ref{fig:noninteracting_line}, show this complementary behavior.

\subsection{\label{sec:hybridsystems} Hybrid Chain-Adatom System}
\begin{figure}
    \centering
    \includegraphics[width=0.48\textwidth]{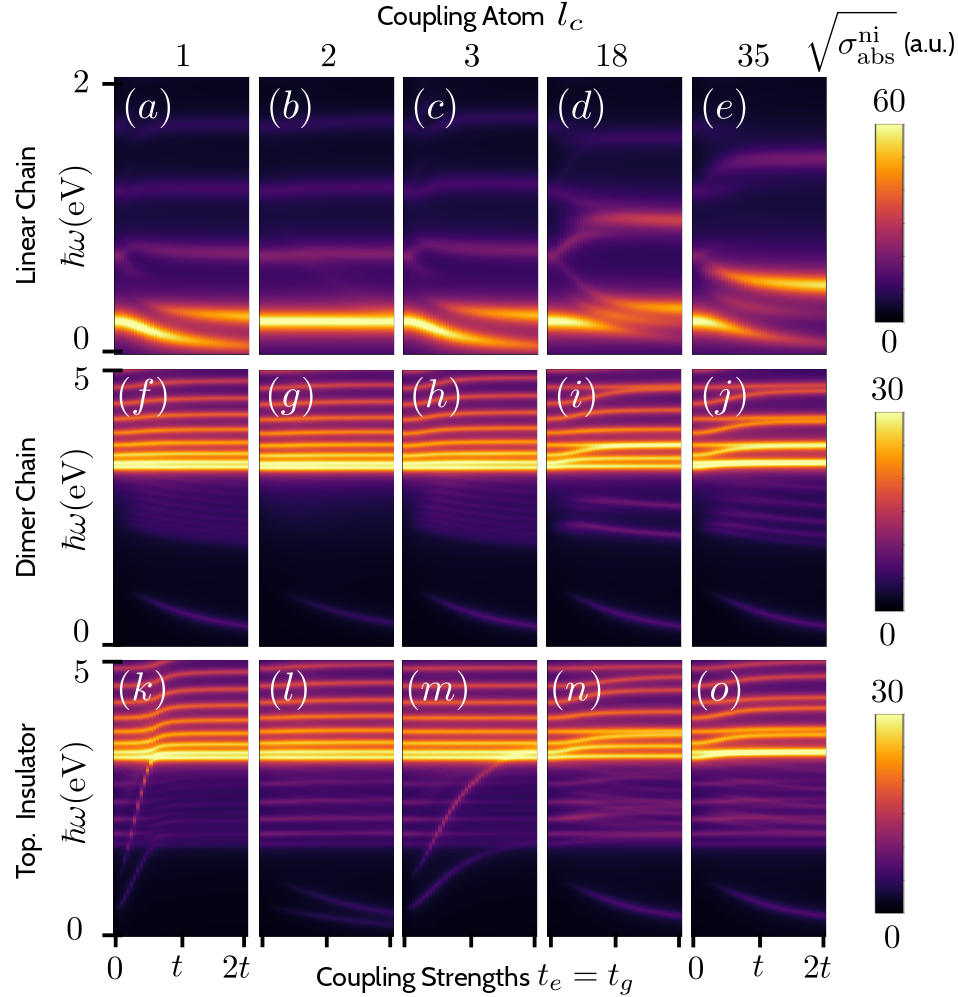}
    \caption{Square root of the \textit{non-interacting} absorption cross sections $\sqrt{\sigma_{\rm abs}^{\rm ni}(\omega)}$ of the linear chain (a-e), dimer chain (f-j), and topological insulator (k-o). The very left column shows the absorption cross section in case the adatom is coupled to the edge atom $l_c=1$ of the chain. The other columns show the same quantity for other coupling positions mentioned in the title of the figure. Please note, that coupling to $l_c=18$ corresponds to partitioning the chain according to the ratio 1:3 and coupling to $l_c=35$ divides the chain in the middle into two parts of equal length. Furthermore, please note that the energy axis of the linear chain is cut at $2\,\rm eV$ whereas the two insulating structures are shown up to $5\,\rm eV$ since they are lacking low-energy modes in the uncoupled case.}
    \label{fig:noninteracting_absorption}
\end{figure}

In the previous sections, we have discussed the electronic and optical properties of the stand-alone chain antennas. In the following, we will discuss the optical absorption in the presence of an adatom when $t_e=t_g>0$. It is instructive to first investigate the case of non-interacting electrons. Within this idealized model one may directly deduce how modifications, that the single-particle states undergo upon sensing the adatom's presence and that are due to the hybridization with the newly introduced adatom states, translate into optical properties through Eq.~\eqref{eq:noninteracting}. In contrast, effects that manifest due to Coulomb interaction can be analyzed in an isolated manner from the previously mentioned aspect and are discussed in Sec.~\ref{sec:interactingsystem}.\\
Figure~\ref{fig:noninteracting_absorption} shows the non-interacting absorption cross section of the hybrid chain-adatom systems as a function of coupling strengths, \textit{i.e.}, varying chain-adatom distances or TLS dipole orientations, and for different chain coupling atoms $l_c$. The evolution of the system as a funtion of the chain-adatom coupling strength is presented in Figs.~\ref{fig:characteristics_chain} and \ref{fig:energy_parity}. The former shows the state activity, the hybridization measure, and the localization of those states of the hybrid linear chain-adatom system that are close to the Fermi energy. The latter depicts the energy landscape of the three chains as a function of chain-adatom coupling strength; the color of the lines encodes the parity of the wavefunction part on the chain of the respective states according to $\mathcal{P}_{|j\rangle}=\langle j|\hat{P}|j\rangle=\sum_{l=1}^{N_A} c_{j,l}c_{j,N_A+1-l}$.\\
In general, we note the linear chain to be much more prone to hybridize with the adatom and change its optical properties than the other systems under consideration. We show the main results for all three SSH model structures. However, we limit our detailed discussion to the more attractive case of the linear chain.

\subsubsection{Coupling to the Edge}
\begin{figure}
    \centering
    \includegraphics[width=0.48\textwidth]{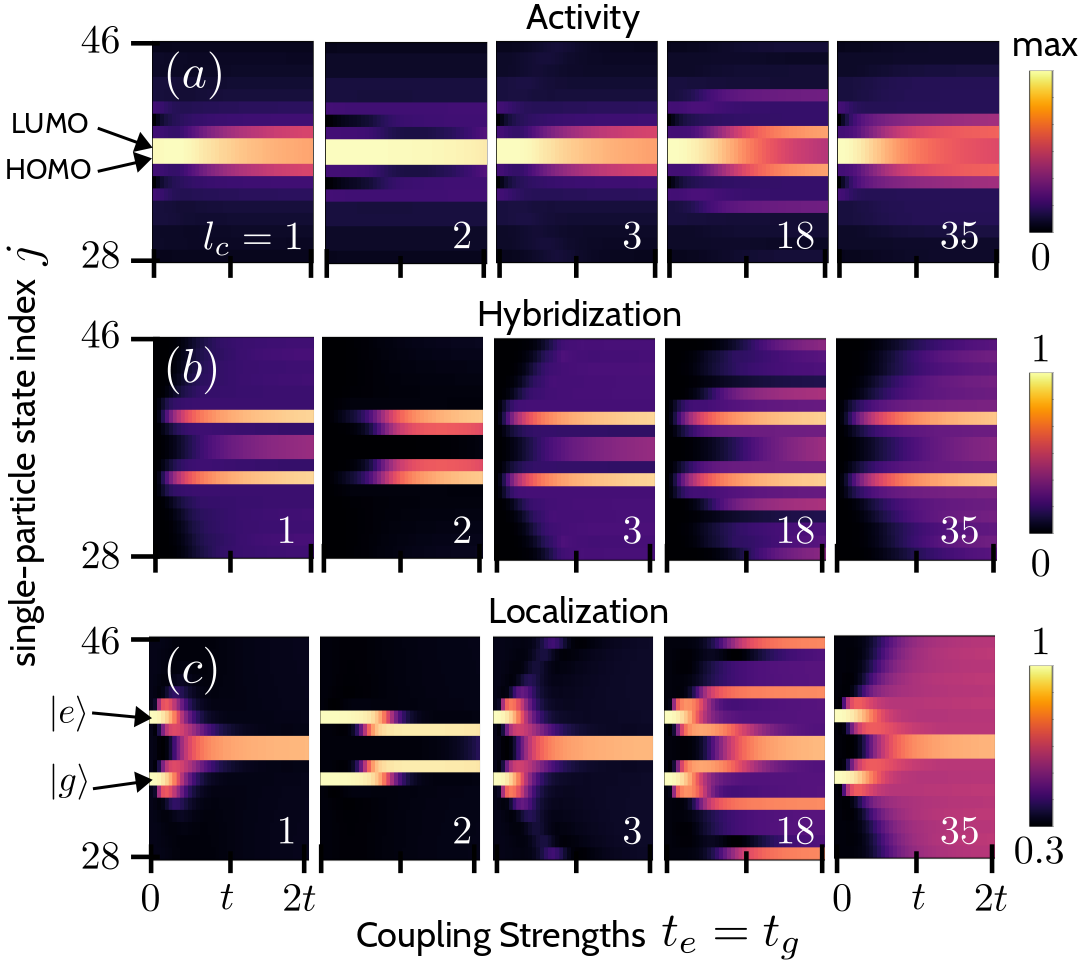}
    \caption{(a) State activity $a_{|j\rangle}$ for states closely below and above the Fermi energy of the linear chain as a function of the coupling strengths $t_e=t_g$ of the adatom states to $l_c$. We investigate different coupling positions $l_c\in\{1, 2, 3, 18, 35\}$. (b) Hybridization $h_{|j\rangle}$ of the adatom states with the states of the linear chain for different coupling positions. (c) Localization of the states of the hybrid linear chain-adatom system.}
    \label{fig:characteristics_chain}
\end{figure}

\begin{figure}
    \centering
    \includegraphics[width=0.48\textwidth]{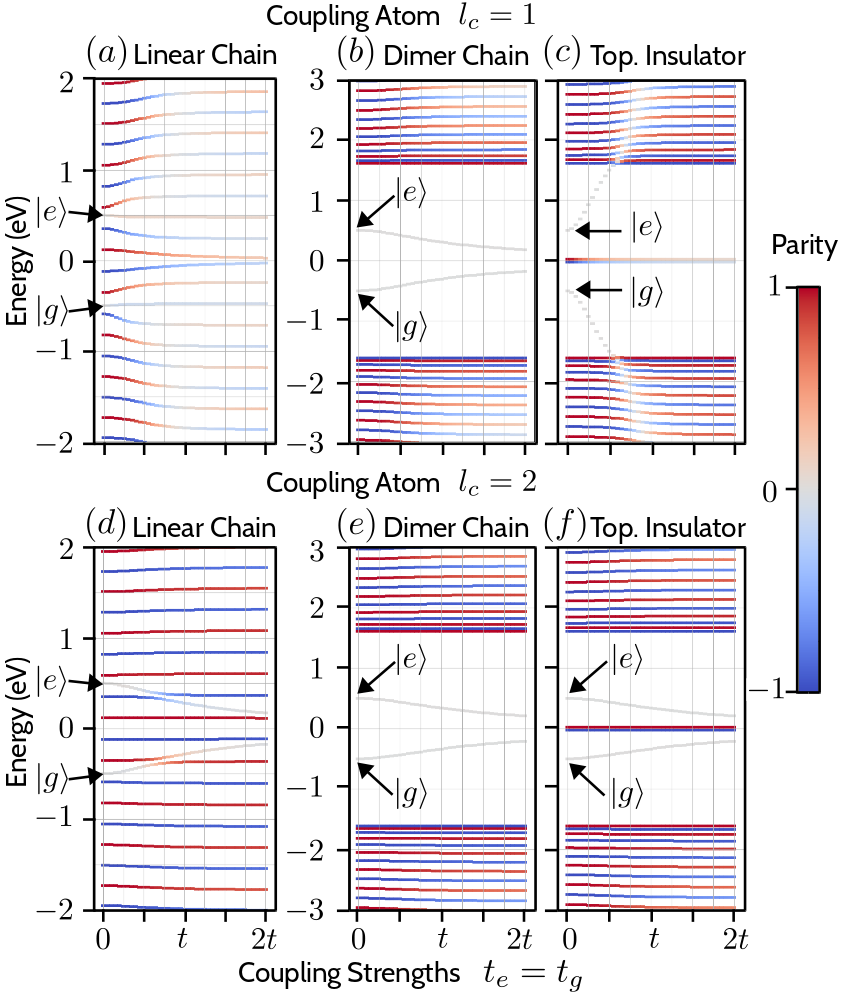}
    \caption{The energy landscape of the linear chain (a,d), the dimer chain (b,e), and the topological insulator (c,f) as a function of coupling strengths $t_e$ and $t_g$ for coupling locations $l_c=1$ (a-c) and $l_c=2$ (d-f). The color indicates the parity $\mathcal{P}_{|j\rangle}=\langle j|\hat{P}|j\rangle$ of the part of the wavefunction that is localized on the chain sites, and which is the discretized analogon of $\langle\psi_j(\mathbf{r})|\psi_j(-\mathbf{r})\rangle$ in our framework. The edge states of the topological insulator have been slightly shifted away from zero for the sake of better visibility.}
    \label{fig:energy_parity}
\end{figure}

The absorption spectra of the hybrid linear chain-adatom system in the top row of Fig.~\ref{fig:noninteracting_absorption} show that the coupling position plays a crucial role for the optical absorption. While coupling to $l_c=1$ and $l_c=3$ shows a similar effect, we notice that the absorption spectrum of the system is barely affected if one couples the adatom to $l_c=2$. This observation can be explained via the absolute value of the real-space expansion coefficients $|c_{j1}|$ and $|c_{j3}|$ of the stand-alone chain's energy eigenstates, which are energetically closest to the adatom's states at $\pm 0.5\,\rm eV$. They are significantly larger than $|c_{j2}|$. In fact, $|c_{j2}|\approx 0$ holds true not only for the HOMO and LUMO states (as can be seen in Fig.~\ref{fig:chains_with_states}a), but also for the other states in the vicinity of the Fermi energy $E=0$. As a consequence, coupling effects are negligible in this configuration.\\
In Figs.~\ref{fig:noninteracting_absorption}a and c, we notice a strong red-shift of the most prominent low-energy HOMO-LUMO transition mode which is accompanied by a decrease in energy difference of the HOMO and LUMO states in the energy landscape, as can be confirmed in Fig.~\ref{fig:energy_parity}a. At the same time, the mode intensity drops for higher coupling strengths, since the HOMO and LUMO states, that were of purely odd and even parity in the uncoupled case, $\mathcal{P}=-1$ and $\mathcal{P}=1$, change their symmetry behavior and couple less strongly. Moreover, another prominent mode builds up in the same spectral region. As can be seen from Fig.~\ref{fig:characteristics_chain}a, especially the states HOMO$-1$ and LUMO$+1$ become optically active, when the adatom is coupled stronger to the linear chain. Indeed, a thorough analysis of this newly occurring mode reveals that it is related to the transitions HOMO$-1\rightarrow$ LUMO and HOMO $\rightarrow$ LUMO$+1$. In Fig.~\ref{fig:energy_parity}a, we see that the transition HOMO$-1$ $\rightarrow$ LUMO is symmetry-forbidden in the uncoupled system, since both states are of even parity $\mathcal{P}=1$. By increasing the coupling strength, however, the transition becomes allowed and manifests itself in Fig.~\ref{fig:noninteracting_absorption} as the previously mentioned mode of increasing intensity.\\


To further illustrate the inertia of the chain to couple to the adatom for $l_c=2$ in Fig.~\ref{fig:noninteracting_absorption}b, we compare the hybridization in Fig.~\ref{fig:characteristics_chain}b for $l_c=1,3$ and $l_c=2$. In the former two cases ($l_c=1,3$) we observe that all states in the given range show non-zero hybridization already for adatom coupling strength below $t$, \textit{i.e.}, they sense the presence of the adatom and are modified accordingly. In Fig.~\ref{fig:energy_parity}a, this is reflected in the fact that all states are spectrally shifted and lose their well-defined symmetry. In the latter case $(l_c=2)$, however, we observe a vanishing hybridization for almost all states. Please note that the hybridization values for the two states below the HOMO and above the LUMO are non-zero only because they interchange their index (see Fig.~\ref{fig:energy_parity}d). This is not the case for $l_c=1,3$ (see Fig.~\ref{fig:energy_parity}a). Furthermore, Fig.~\ref{fig:energy_parity}d reveals that the spectrum is not modified much by the adatom states. In particular, we notice that all states retain their parity and only states HOMO-1 and LUMO+1 interact with the adatom when they change their index in an anticrossing pattern.\\
Figure~\ref{fig:characteristics_chain}c shows the localization of the hybrid linear chain-adatom system's states around the Fermi energy. In the uncoupled case ($t_e=t_g=0)$, we observe that the adatom's states $|e\rangle$ and $|g\rangle$ exhibit a localization of 1, since the state's real-space wavefunction is fully localized on the respective orbital of the adatom. All other states attain a localization value of $L_{\rm crit}\approx 0.33$, as already discussed in Sec.~\ref{sec:linear_chain}. Again, we observe a qualitatively different behavior of the localization for $l_c=1,3$ on the one hand and for $l_c=2$ on the other hand. In the latter case, the ground and excited states of the adatom interchange indices around $t_e=t_g\approx 0.8t$ in the energy level diagram with the states HOMO$-1$ and LUMO$+1$, respectively (see also Fig.~\ref{fig:energy_parity}d). The states attributed to the linear chain's continuum barely change their energy as a function of coupling strength. For $l_c=1,3$, on the other hand, the energy landscape does not exhibit energy level crossings, but the adatom's states fit seamlessly into the state continuum of the linear chain. An almost equally spaced energy ladder is building up again, similar to the energy level diagram in Fig.~\ref{fig:chains_with_states}a, however, incorporating the adatom's orbitals (Fig.~\ref{fig:energy_parity}a).\\
Figures \ref{fig:energy_parity}b and e show the energy landscape of the dimer chain for $l_c=1,2$ as a function of coupling strength. We note that the adatom states in the energy gap of the insulator approach each other and produce a small-intensity low-energetic and red-shifting mode in the non-interacting absorption cross section in Figs.~\ref{fig:noninteracting_absorption}f and g. The remaining spectrum remains mostly unmodified.\\
The topological insulator's absorption cross section in Fig.~\ref{fig:noninteracting_absorption}k exhibits two strongly blue-shifting modes for comparably small coupling strengths already. From Fig.~\ref{fig:energy_parity}c we deduce that the higher-energetic one belongs to the transition between the strongly dispersive parity-less modes that dive into the quasicontinua at a coupling strength around $0.5t$. The lower-energetic mode can be related to the transition from one of the strongly dispersive parity-less states to the edge states. Hence, their energies exactly differ by a factor of two. As revealed by Fig.~\ref{fig:energy_parity}f, the topological insulator is less reactive for $l_c=2$. The only modifications we see in Fig.~\ref{fig:noninteracting_absorption}l is the build-up of two red-shifting weak modes in the low-energy region related to the states within the band gap, reminiscent of the dimer chain.

\begin{figure*}
    \centering
    \includegraphics[width=0.9\textwidth]{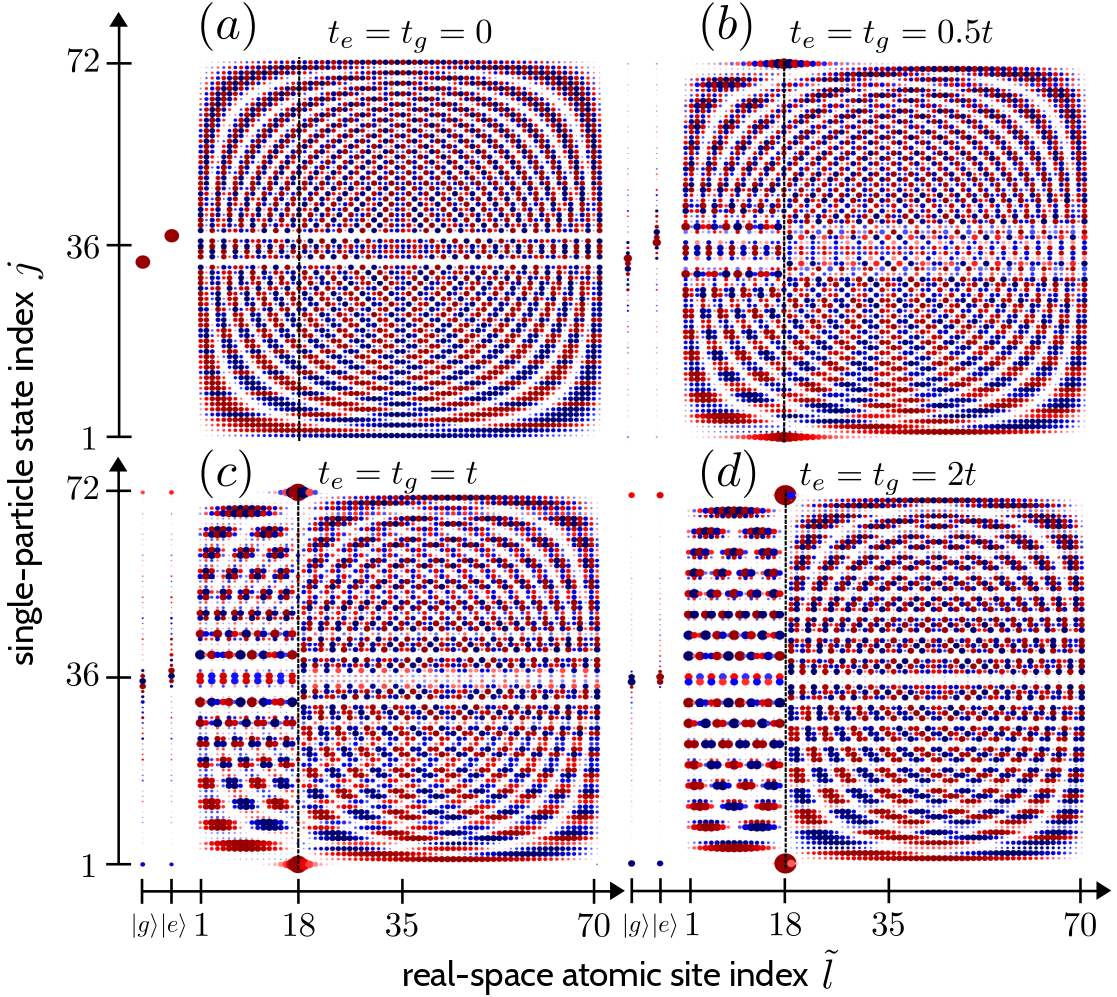}
    \caption{Real-space representations of the single-particle energy states of the hybrid linear chain-adatom system for different coupling strengths (a) $t_e=t_g=0$, (b) $t_e=t_g=0.5t$, (c) $t_e=t_g=t$, and (d) $t_e=t_g=2t$. The adatom is coupled to the host atom $l_c=18$ in all subfigures (indicated by the black dashed line); the adatom's states $|e\rangle$ and $|g\rangle$ are depicted to the left of the chain continuum. Like in Fig.~\ref{fig:chains_with_states}, the circles' color encodes the real-valued expansion coefficients $c_{jl}$, and their size is proportional to $|c_{jl}|^2$.}
    \label{fig:states_coupling}
\end{figure*}

\subsubsection{Coupling to the Bulk}
Before discussing the two right columns of Fig.~\ref{fig:characteristics_chain} where we couple the adatom to bulk sites $l_c=18$ and $l_c=35$ of the chain, we need to understand Fig.~\ref{fig:states_coupling} first. It shows the real-space representations of the single-particle energy states of the hybrid linear chain-adatom system for different coupling strengths $t_e=t_g=0$ (a), $t_e=t_g=0.5t$ (b), $t_e=t_g=t$ (c), $t_e=t_g=2t$ (d). We have always coupled the adatom to $l_c=18$, which divides the chain geometrically according to the ratio 1:3. In the decoupled system (a), the linear chain and the adatom are not hybridized and the real-space wavefunction either lives completely on the adatom orbitals ($|e\rangle$ and $|g\rangle$) or completely on the chain (all other states).  When we increase the coupling (decrease the distance of the adatom to the chain or align its dipole moment suitably), we observe in Fig.~\ref{fig:states_coupling}b that we induce population on the adatom's sites for a significant number of energy eigenstates. Simultaneously, the wavefunctions of the lowest-energy and highest-energy states get attracted by the adatom. By further increasing the coupling strength (Figs.~\ref{fig:states_coupling}c and d), we observe that the adatom acts as a potential barrier for the wavefunction and effectively splits the chain apart into two stand-alone chains of smaller lengths. The real-space wavefunctions of most of the energy eigenstates are apparently locked on either side of the chain. Exceptions thereof are i) the lowest-energy $(j=1)$ and highest-energy $(j=72)$ modes which are strongly localized on the adatom and in the close vicinity of the coupling atom, and ii) the HOMO and LUMO of the strongly coupled system which are localized on the adatom and on the shorter part of the chain. It is interesting to observe that the single-particle electronic structure of the whole hybrid system seemingly collapses into a small chain on the left of the adatom in Fig.~\ref{fig:states_coupling} and a longer part on the right. The 18 atoms belonging to the smaller sub-part of the chain host 9 prominent energy eigenstates below the Fermi energy and 9 above. This sums up to 18 states, which is exactly the expected structure for a stand-alone linear chain of 18 atoms. The larger part of the chain behaves accordingly. Especially in the vicinity of the Fermi energy where the optically active states are hosted, we notice that every fourth state is localized on the left shorter side of the system, reflecting the partitioning ratio of the chain. These geometrical features are also apparent in the localization figure of merit in Fig.~\ref{fig:characteristics_chain}c for $l_c=18$, where we see that one in four states shows a substantially increased localization for high coupling strengths, when the real-space wavefunction is localized on only 18 atoms. Moreover, we notice that the adatom population on the optically active states in the region around the Fermi energy is maximum for intermediate couplings rather than for larger or smaller ones.\\
In the non-interacting absorption spectrum of this configuration shown in Fig.~\ref{fig:noninteracting_absorption}d, this translates to the emergence of two prominent modes for high coupling strengths. We observe that the resonant mode at $0.24\,\rm eV$ in the uncoupled case evolves toward two modes at $0.33\,\rm eV$ and $0.98\,\rm eV$. While the former energy coincides with the HOMO-LUMO gap of a stand-alone chain made up by 52 atoms, the latter is equivalent to the HOMO-LUMO gap of an 18-atomic chain. 
Equivalently, Fig.~\ref{fig:noninteracting_absorption}e shows the spectrum of the configuration where the chain is split close to the middle. Effectively, the 70-atomic chain with HOMO-LUMO gap of $0.24 \,\rm eV$ collapses into two 35-atomic chains with HOMO-LUMO gaps of $0.49\,\rm eV$, which results in the appearance of a mode at $0.49\,\rm eV$ for high coupling strengths. Moreover, Fig.~\ref{fig:characteristics_chain}c for $l_c=35$ further reveals, that besides the HOMO and LUMO states which have a significant share of population concentrated on the adatom for high coupling strengths, all other states attain a similar localization value, just as it was the case for the stand-alone linear chain. In this case, however, the wavefunctions are localized on one of the two almost equally long sides of the chain. This leads to localization values around $L_{|j\rangle}\approx 0.67$.

\subsection{\label{sec:interactingsystem} Interacting Hybrid System}
\begin{figure}
    \centering
    \includegraphics[width=0.48\textwidth]{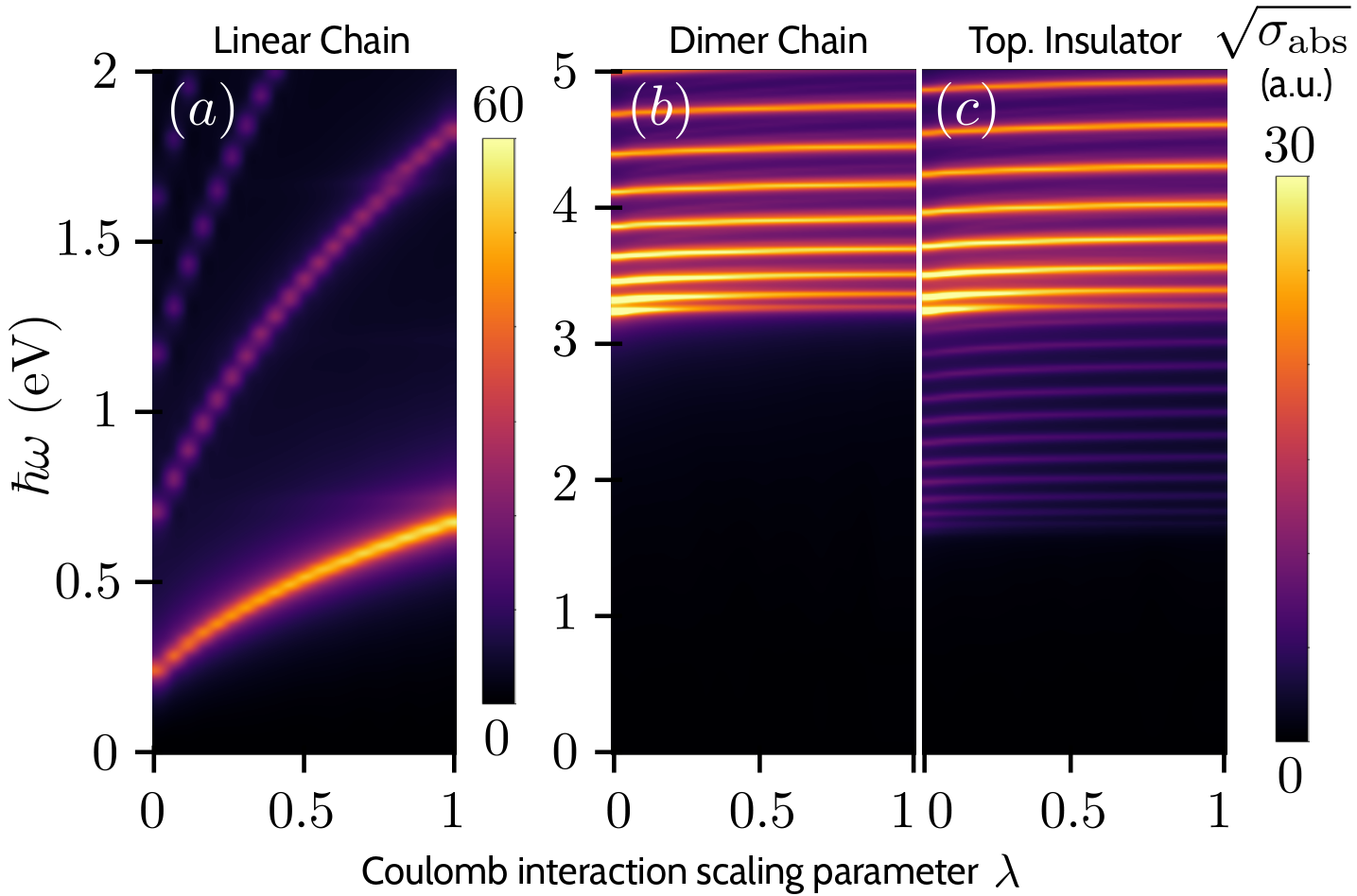}
    \caption{Square root of the linear absorption cross sections of (a) the linear chain, (b) the dimer chain, and (c) the topological insulator as a function of the Coulomb interaction scaling parameter $\lambda$. For $\lambda=0$, we obtain the non-interacting absorption cross section $\sigma^{\rm ni}_{\rm abs}$ analyzed in Fig.~\ref{fig:noninteracting_absorption}; $\lambda=1$ corresponds to the fully interacting systems shown in Fig.~\ref{fig:interacting_absorption}. We find that electron-electron interaction heavily modifies the absorption characteristics of the linear chain, whereas the dimer chain and the topological insulator are barely influenced by Coulomb interaction. For the latter two insulating structures, the non-interacting absorption cross section is a valid approximation to the fully interacting system.}
    \label{fig:coulomb_scaling}
\end{figure}
So far, we have considered non-interacting electrons. In insulating electronic systems whose optical properties are predominantly characterized by single-particle transitions, Coulomb interaction may often be safely neglected~\cite{Bernadotte2013}. However, previous contributions have revealed that especially in the metallic linear chain, Coulomb interaction leading to collective plasmonic charge carrier oscillations plays a significant role for determining resonant modes~\cite{Bernadotte2013, Mueller2020, deVega2020, Townsend2015,Gao2005,Yan2007,Townsend2021}. 
To elucidate this, we show the absorption spectra of all three stand-alone structures in Fig.~\ref{fig:coulomb_scaling} as a function of the Coulomb interaction strength $\lambda$ introduced in App.~\ref{app:time_dependent}.\\
In Fig.~\ref{fig:coulomb_scaling}a, we verify that the linear chain's optical modes strongly depend on the Coulomb interaction strength in a continuous way. The lowest-energy most prominent mode shifts from $\hbar\omega=0.24\,\rm eV$ for $\lambda=0$ (Coulomb interaction turned off) to $0.67\,\rm eV$ for $\lambda=1$ (Coulomb interaction fully taken into account). The next higher mode even changes its spectral position by more than $1\,\rm eV$ upon turning on electron-electron interaction. We conclude that to properly find resonant modes of the combined linear chain-adatom system, performing interacting simulations which take into account Coulomb repulsion is inevitable for the hybrid system as well. However, Fig.~\ref{fig:coulomb_scaling}a also reveals that the transition from the lowest-energy mode of the \textit{non-interacting} system to the corresponding mode of the \textit{interacting} system is smooth and continuous. It can therefore be concluded that the HOMO and LUMO continue to constitute the predominantly involved single-particle states in the formation of this resonance also in the interacting system. More detailed investigations concerning this issue have been performed and similar conclusions have been drawn in the context of non-interacting and interacting molecular chains in a TB framework~\cite{deVega2020}, in time-dependent density functional theory~\cite{Bernadotte2013}, and upon exact diagonalization~\cite{Townsend2015}, as well as in metallic gold nanospheres~\cite{Townsend2015, Townsend2012}, and in structured nanographene~\cite{Mueller2021}.\\
\begin{figure}
    \centering
    \includegraphics[width=0.48\textwidth]{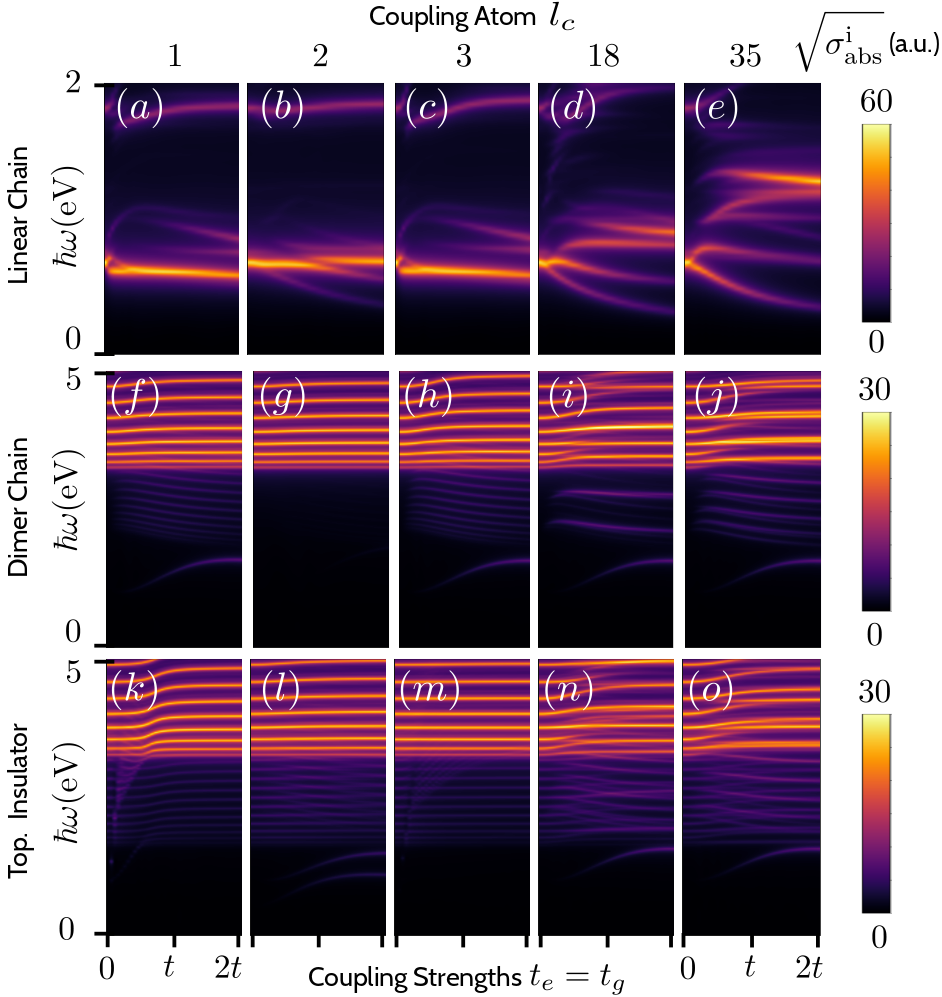}
    \caption{Square root of the \textit{interacting} absorption cross sections $\sqrt{\sigma_{\rm abs}^{\rm i}(\omega)}$ of the linear chain (a-e), the dimer chain (f-j), and the topological insulator (k-o). The very left column shows the absorption cross section in case the adatom is coupled to atom $l_c=1$ of the chain. The other columns show the same quantity for other coupling positions.}
    \label{fig:interacting_absorption}
\end{figure}
On the other hand, Figs.~\ref{fig:coulomb_scaling}b and c reveal that the absorption spectra of the dimer chain and the topological insulator hardly depend on the Coulomb interaction scaling parameter $\lambda$, which renders their \textit{non-interacting} absorption spectra in Fig.~\ref{fig:noninteracting_absorption} more reliable than that of the linear chain. This statement holds true even in the presence of the adatom as can be seen by comparing the second and third rows of Figs.~\ref{fig:noninteracting_absorption} and \ref{fig:interacting_absorption}.\\
Figure~\ref{fig:interacting_absorption} shows the \textit{interacting} absorption spectrum of the linear chain (a-e), the dimer chain (f-j), and the topological insulator (k-o) as a function of the coupling strengths $t_e$ and $t_g$ for different coupling atom positions $l_c$. The dimer chain's and the topological insulator's absorption spectra exhibit only minor changes upon increasing the coupling strengths. Small modifications can be perceived, however, especially when coupling the adatom to a bulk atom, \textit{i.e.}, when $l_c=18$ or $l_c=35$. Then, the modes in the quasi-continua above the band gap become more pronounced but their number decreases.\\
In contrast, the linear chain's absorption spectra show a rich variety of spectral features already for small coupling strengths. In case of coupling the adatom to atoms at the edge (Figs.~\ref{fig:interacting_absorption}a-c), the mode at $\hbar\omega=0.67\,\rm eV$ persists relatively stable as a function of coupling strength and is the only remaining prominent one from Figs.~\ref{fig:noninteracting_absorption}a-c. Coupling the adatom to the bulk atoms (Figs.~\ref{fig:interacting_absorption}d and e), however, paves the way for manipulating the modes in the range between $0.35\,\rm eV$ and $1\,\rm eV$ for $l_c=18$ (Fig.~\ref{fig:interacting_absorption}d), and between $0.35\,\rm eV$ and $1.5\,\rm eV$ for $l_c=35$ (Fig.~\ref{fig:interacting_absorption}e). Consistent with Fig.~\ref{fig:coulomb_scaling}a, the apparent modes are blue-shifted with respect to the non-interacting systems.

\section{\label{sec:summary} Summary}
In this paper, we have investigated the tunability of the optical modes of one-dimensional atomic chains upon coupling them to adatom impurities. To that end, we have presented a tight-binding based hybrid system model for an atomic SSH chain, applicable to linear organic molecules such as polyenes, that is coupled to an adatom treated as a two-level system. We investigated in detail how the adatom influences the real-space representations of the model Hamiltonian's single-particle eigenstates and how the optical absorption cross section is modified as a function of coupling strength between the adatom and the chains. We have shown that in certain coupling positions the adatom may significantly modify the optical properties especially of the metallic linear chain antenna already for comparably small coupling strengths. We conclude that at high coupling strengths, the adatom acts as a potential barrier and effectively splits the chain apart into two sub-systems. For suitable parameter sets, this is reflected as well in the absorption spectrum which no longer shows one single pronounced low-energy mode, but two higher-energy modes that stem from the states of two smaller chains. We find that the dimerized chain and the topological insulator described within the Su-Schrieffer-Heeger model remain both relatively inert to the presence of the adatom. The linear chain's modes, however, can be readily tuned over a broad spectral region by changing the interaction strengths that couple the adatom to the chains, \textit{i.e.}, the distance of the adatom to the chain or the orientation of its dipole moment.

\begin{acknowledgments}
M.M.M. acknowledges financial support through the Research Travel Grant by the Karlsruhe House of Young Scientists (KHYS). M.M.M. and C.R. acknowledge support by the Deutsche Forschungsgemeinschaft (DFG, German Research Foundation) (Project No. 378579271) within Project RO 3640/8-1 and from the VolkswagenStiftung. M.K. and K.S. acknowledge the support from the National Science Centre, Poland (Project No. 2016/23/G/ST3/04045). A.A. acknowledges the Spanish Ministry of Science and Innovation with  
PID2019-105488GB-I00 and PCI2019-103657, the Basque Government through  
the University of the Basque Country Project No. IT-1246- 19, the  
European Commission from the NRG-STORAGE Project (No. GA 870114) and  
H2020-FET OPEN Project MIRACLE (No. GA 964450).
\end{acknowledgments}

\appendix

\section{\label{app:time_dependent} Computation of the time-dependent dipole moment $\mathbf{p}(t)$}
The Hamiltonian of the hybrid system which is coupled to a time-dependent electric field $\mathbf{E}(t)$ reads 
\begin{align}
    H(t) &= H_{\rm antenna} + H_{\rm TLS} + H_{\rm interaction}-e\varphi(t),
    \label{eq:AppA}
\end{align}
where the time-independent part is given by Eq.~\eqref{eq:Hsystem}, $e$ is the electronic charge, and $\varphi(t)=\varphi^{\rm ext}(t)+\varphi^{\rm ind}(t)$ is the total electric potential, composed of the externally applied potential $\varphi^{\rm ext}_l(t)=-\mathbf{r}_l\cdot\mathbf{E}(t)$ and the induced potential $\varphi^{\rm ind}_{\tilde{l}}(t)=-\lambda eN_e\sum_{\tilde{l'}}v^{\rm hyb}_{\tilde{l}\tilde{l}'}\left(\rho_{\tilde{l}'\tilde{l}'}(t)-\frac{1}{N}\right)$ at orbital $\tilde{l}$. Here, $N_e$ is the number of electrons in the system and $v^{\rm hyb}_{\tilde{l}\tilde{l}'}$ denotes the Coulomb interaction matrix element which couples the charges on sites $\tilde{l}$ and $\tilde{l}'$. The parameter $\lambda\in[0,1]$ has been introduced to artificially turn on and off the Coulomb interaction in the simulations. The elements of the Coulomb matrix that couple antenna orbitals to other antenna orbitals are contructed as follows: For the onsite-values, nearest neighbor-, and next-to-nearest neighbor-values, we apply values calculated by Potasz \textit{et. al.}~\cite{Potasz2010}; the values coupling sites further away from each other follow the $1/r$ Coulomb power law. We outline the construction of the adatom's contributions to the Coulomb matrix $v^{\rm hyb}$ in App.~\ref{app:hybrid_coulomb}. The density matrix $\rho$ is initialized as $\rho^0 = \frac{2}{N}\sum_{j}f_j|j\rangle\langle j|$, where $f_j$ is the occupation of the energy eigenstate $|j\rangle$ of the Hamiltonian given in Eq.~\eqref{eq:AppA} without externally applied electric field, and the factor 2 in the numerator accounts for spin degeneracy. The density matrix is propagated through time according to the master equation
\begin{align}
    \frac{\partial}{\partial t}\rho(t) = -\frac{i}{\hbar}[H(t),\rho(t)]-\frac{1}{2\tau}\left(\rho(t)-\rho^0\right).
\end{align}
Here, $\hbar\tau^{-1}=10\,\rm meV$ is a phenomenological scattering energy introduced to mimic dissipation. The value is taken from doped extended bulk graphene~\cite{Cox2014}. We determine the resulting dipole moment $\mathbf{p}(t)=\sum_{\tilde{l}}\mathbf{r}_{\tilde{l}} q_{\tilde{l}}(t)$, where $q_l(t)=-eN_e\rho_{ll}(t)$ is the charge at site $l$ as a function of time.

\section{\label{app:hybrid_coulomb} Computation of the inter-system values of the total Coulomb matrix $v^{\rm hyb}$}
The extended Coulomb interaction matrix $v^{\rm hyb}$ for the hybrid system is based on the interaction matrix of the stand-alone antenna $v$. The latter is determined as described in previous contributions~\cite{Cox2014, Mueller2020}. The construction of $v^{\rm hyb}$ is as follows:
\begin{align}
    v^{\rm hyb}_{ll'} &= v_{ll'}\qquad \text{ for } l,l'\in[1,N_a].
    \label{eq:intraantenna}
\end{align}
Like on the diagonal of $v$, we impose the on-site values $v_{\rm os}=16.52\,\rm eV$ on the part of $v^{\rm hyb}$ which corresponds to the adatom sites, $v^{\rm hyb}_{gg} = v^{\rm hyb}_{ee}= v^{\rm hyb}_{eg}= v^{\rm hyb}_{ge}= v_{\rm os}$. The inter-system elements which couple the antenna and the adatom are determined according to
\begin{align}
    v^{\rm hyb}_{el} = v^{\rm hyb}_{le} &= v_{ll_c}\cdot \frac{a_{cc}}{d+a_{cc}}\sqrt{\left|\frac{t_e}{t}\right|},\\
    v^{\rm hyb}_{gl} = v^{\rm hyb}_{lg} &= v_{ll_c}\cdot \frac{a_{cc}}{d+a_{cc}}\sqrt{\left|\frac{t_g}{t}\right|},\\
    v^{\rm hyb}_{el_c} = v^{\rm hyb}_{l_ce} &= v_{\rm nn} \cdot \frac{a_{cc}}{d}\sqrt{\left|\frac{t_e}{t}\right|},\\
    v^{\rm hyb}_{gl_c} = v^{\rm hyb}_{l_cg} &= v_{\rm nn} \cdot \frac{a_{cc}}{d}\sqrt{\left|\frac{t_g}{t}\right|},
\end{align}
where $l\in [1,N_a]\backslash \{l_c\}$ and $v_{\rm nn}=8.64\,\rm eV$ is the experimentally determined nearest-neighbor value for honeycomb carbon~\cite{Potasz2010}. $d$ is the distance of the adatom to the coupling atom $l_c$ and $a_{\rm cc}=1.42\,\rm \AA$ is the distance of two carbon atoms in the antenna. The distance $d$ relates to the couplings according to $t_{e,g}=t\left(\frac{a_{cc}}{d}\right)^2$ \cite{Harrison1977} and the intra-antenna values $v_{ll_c}$ are determined through Eq.~\eqref{eq:intraantenna}.

\providecommand{\noopsort}[1]{}\providecommand{\singleletter}[1]{#1}%

\end{document}